\documentclass[useAMS,usenatbib]{mn2e}
\usepackage{graphicx}
\def\aj{AJ}%
\def\aap{A\&A}%
\def\nat{Nature}%
\usepackage{times}
\title[Probing a massive radio galaxy with gravitational lensing]{Probing a massive radio galaxy with gravitational lensing}
\author[A. More et al.]{A. More,$^{1}$\thanks{E-mail: anupreeta@mpifr-bonn.mpg.de} J. P. McKean,$^{1}$ T. W. B. Muxlow,$^2$ R. W. Porcas,$^{1}$  C. D. Fassnacht$^{3}$ and 
\newauthor L. V. E. Koopmans$^{4}$\\
$^{1}$Max-Planck-Institut f\"{u}r Radioastronomie, Auf dem H\"{u}gel 69, D-53121 Bonn, Germany\\
$^{2}$MERLIN/VLBI National Facility, Jodrell Bank Observatory, University of Manchester, Macclesfield, Cheshire SK11 9DL\\
$^{3}$Department of Physics, University of California, Davis, CA 5616, USA\\
$^{4}$Kapteyn Astronomical Institute, Postbus 800, NL-9700 AV Groningen, the Netherlands}

\begin{document}

\date{Accepted 2007 December 6. Received 2007 December 5; in original form 2007 October 29}

\pagerange{\pageref{firstpage}--\pageref{lastpage}} \pubyear{2002}

\maketitle

\label{firstpage}

\begin{abstract}
The gravitational lens system CLASS B2108+213 has two lensed images separated by 4.56 arcsec. Such a wide image separation suggests that the lens is either a massive galaxy, or is composed of a group of galaxies. To investigate the structure of the lensing potential we have carried out new high resolution imaging of the two lensed images at 1.7~GHz with the Very Long Baseline Array (VLBA) and at 5~GHz with global Very Long Baseline Interferometry (VLBI). Compact and extended emission is detected from the two lensed images, which provides additional constraints to the lensing mass model. We find that the data are consistent with either a single lensing galaxy, or a two galaxy lens model that takes account of a nearby companion to the main lensing galaxy within the Einstein radius of the system. However, for an ensemble of global power-law mass models, those with density profiles steeper than isothermal are a better fit. The best-fitting profile for a single spherical mass model has a slope of $\gamma=$~2.45$_{-0.18}^{+0.19}$ . The system also has a third radio component which is coincident with the main lensing galaxy. This component is detected at milli-arcsecond scales for the first time by the 1.7~GHz VLBA and 5~GHz global VLBI imaging. However, the third radio component is found not to be consistent with a core lensed image because the radio spectrum differs from the two lensed images, and its flux-density is too high when compared to what is expected from simple mass models with a variable power-law density profile and/or a reasonable core radius. Furthermore, 1.4~GHz imaging of the system with the Multi-Element Radio Link Interferometric Network (MERLIN) finds extended lobe emission on either side of the main lensing galaxy. Therefore, the radio emission from the third radio component is almost certainly from an AGN within the main lensing galaxy, which is classified as an FR\,I type radio source.
\end{abstract}


\section{Introduction}

Gravitational lensing is caused by the deflection of light from a distant object by matter along the line-of-sight to the observer. If the surface mass density of an intervening matter distribution is above some critical density then multiple images of the background source are formed. The positions and flux-densities of these lensed images can provide a powerful probe of the enclosed mass and shape of the lensing matter distribution (e.g. \citealt*{schneider92}). If this mass constraint is combined with an additional estimate of the mass at another radius, then the average matter density slope [$\gamma$, where the matter density $\rho(r)\propto r^{-\gamma}$] of the combined luminous and dark matter distribution can be established (e.g. \citealt{sand05,koopmans06,gavazzi07}). For galaxy-scale haloes in the mass range 10$^{11}$--10$^{12}$~M$_{\odot}$, the results from combined lensing and stellar kinematic data suggest that the inner 0.3--5 effective radii of galaxies are consistent with an isothermal mass distribution ($\gamma =$~2.01$^{+0.02}_{-0.03}$; \citealt{treu04,koopmans06}).

There are now over 100 examples of galaxy-scale lensing known. These systems have been found from systematic searches around potential lens galaxies (e.g. \citealt{fassnacht04,bolton06,moustakas07}) and from large surveys of the lensed source parent population (e.g. \citealt{bahcall92,king99,browne03,myers03}). Serendipitous discoveries of gravitational lensing from deep high-resolution optical and infrared imaging have also been made (e.g. \citealt{fassnacht06b}). Typically, the image separations for these gravitational lens systems are between 0.5--1.5~arcsec, with only a handful having image separations $\ga$~3~arcsec \citep{lawrence84,wisotzki93,munoz01,sluse03,oguri04,mckean05,bolton06a,inada06}. Since the enclosed mass of a lens system is proportional to the square of the image separation (i.e. $M_E \propto \Delta\theta^2$ for a circularly symmetric mass distribution; \citealt{kochanek91}), such wide image separation lens systems (i.e. $\ga$~3~arcsec) may be due to haloes which are an order of magnitude more massive than those of typical lens galaxies. As such, wide image separation lens systems could be used to probe the matter distribution at the top end of the mass function for galaxy-scale structures. Alternatively, the wide image separation could be due to the lens galaxy being in an over-dense environment, for example, in a group or cluster of galaxies. Recent imaging and spectroscopic surveys of the local environments of lens galaxies have found many to be members of larger structures \citep{fassnacht02,fassnacht06a,momcheva06,williams06,auger07a}. However, the total contribution of the group or cluster to the lensing mass distribution is thought to be no more than $\sim$5 per cent for systems with image separations of $\sim$ 1 arcsecond \citep{momcheva06,auger07a}. Studies of gravitational lens systems with larger image separations may show an enhanced lensing mass distribution which has been boosted by the environment. An extreme example of this is cluster lensing which can produce image separations much larger than 10~arcsec (e.g. \citealt{oguri04,inada06}).

The gravitational lens system CLASS B2108+213 has two radio-loud lensed images separated by 4.56~arcsec \citep{mckean05}. This image separation immediately identifies B2108+213 as an excellent opportunity to study a mass regime between the typical galaxy and cluster-scales. The system was discovered from 8.46~GHz imaging with the Very Large Array (VLA) during the Cosmic Lens All-Sky Survey (CLASS; \citealt{myers03,browne03}). Follow-up observations with the Multi-Element Radio Link Interferometric Network (MERLIN) at 5~GHz found the radio spectra of the two lensed images (A and B) to be similar ($\alpha \sim$~0.15 between 5 and 8.46~GHz, where $S_{\nu} \propto \nu^{\alpha}$), with a flux ratio of $S_B/S_A \sim$~0.5. High resolution radio imaging with the Very Long Baseline Array (VLBA) at 5~GHz found the surface brightness of images A and B to be consistent
with gravitational lensing. Not surprisingly, optical and infrared imaging found a massive elliptical galaxy (G1) at the expected position of the lens. Furthermore, a companion galaxy (G2) was also found within the Einstein radius. A mass model for the system was proposed by \citet{mckean05} which incorporated both G1 and G2 as singular isothermal spheres with an external shear component. This model successfully reproduced the observed positions and flux ratios of the lensed images and required only a small contribution from the field (the external shear was $\sim$2 per cent). However, given the limited number of observational constraints that were provided from the two lensed images, this model has no degrees of freedom. A spectroscopic survey of the environment around the main lensing galaxy has found at least 40 group/cluster galaxies at the same redshift as G1 ($z =$~0.365). Also, the central stellar velocity dispersion of G1 was found to be 360~km\,s$^{-1}$ (McKean et al. in preparation). These new data confirm that G1 is a massive galaxy in a dense group/cluster environment.

Intriguingly, the radio imaging of B2108+213 with the VLA and MERLIN also detected emission from a third radio component (C) which is coincident with the lensing galaxy G1. Lens theory predicts that an extended mass distribution should produce an additional lensed image near the centre of the lens potential (e.g. \citealt{rusin01,keeton03}). The detection of such core lensed images is extremely rare for galaxy-scale systems because the image magnification tends to zero as the inner density profile approaches isothermal. However, the non-detection of core lensed images can place a very strong lower bound to the density profile of the lensing mass distribution (e.g. \citealt{rusin01,keeton03,boyce06,zhang07}). There is currently only one accepted detection of a core lensed image, PMN~J1632$-$0033, whose lensing galaxy has a power law slope of $\gamma \sim$~1.91$\pm$0.02 (\citealt{winn02}; \citealt*{winn03a,winn03b}). The nature of the third radio component of B2108+213 is not entirely clear. The radio spectrum of component C from 5 to 8.46~GHz is consistent with that of the lensed images, A and B. However, the flux-density of component C appears to be too large to be the core lensed image; the flux ratio is $S_C/S_A \sim$~0.1. Furthermore, component C was not detected with the VLBA which suggests that the surface brightness is not consistent with a core lensed image \citep{mckean05}. 

In this paper, we present new high-resolution radio imaging of B2108+213 with the twin aims of determining the nature of the third radio component and finding additional observational constraints for the lensing mass model from the two lensed images. In Section \ref{radio} we show 1.4~GHz imaging of the system with MERLIN which suggests that the emission from component C is from an AGN within the lensing galaxy G1. High resolution imaging of the lensed images at 1.7~GHz with the VLBA and at 5~GHz with global VLBI (Very Long Baseline Interferometry) are also presented in Section \ref{radio}. These new data show extended jet emission from both lensed images on mas-scales. We test gravitational lens mass models for the B2108+213 system in Section \ref{model}. We discuss our results in Section \ref{disc} and present our conclusions in Section \ref{conc}.

Throughout, we assume a $\Omega_M=$~0.3 and $\Omega_{\Lambda}=$~0.7 flat-universe, with a Hubble constant of $H_0=$~70~km\,s$^{-1}$~Mpc$^{-1}$.

\section{Radio Observations}
\label{radio}
In this section, radio imaging of B2108+213 with MERLIN, the VLBA and global VLBI are presented.

\subsection{MERLIN 1.4~GHz observations}
B2108+213 was observed with MERLIN at 1.4~GHz in two runs. The first set of observations were carried out for 14 h on 2005 March 19 and 20 with all of the MERLIN antennas except for the Lovell telescope. The second observing run, which included all of the antennas, lasted for 14 and 8 h on 2005 April 15 and 16, respectively. 3C286 and OQ208 were used as the flux-density and polarization calibrators. A switching cycle of $\sim$2 and $\sim$6 min was used between the phase calibrator (B2103+213) and the lens system. The data were taken in a single IF and divided into 15 channels of 1~MHz width each. The data were taken with both right and left hand circular polarizations. The MERLIN pipeline was used for initial calibration and editing of the data within AIPS\footnote{AIPS - Astronomical Image Processing Software.}.

The MERLIN imaging detected a combination of compact and very low surface brightness extended emission. The total intensity image was initially cleaned with the AIPS task \textsc{imagr} to subtract the higher surface brightness compact radio components (A, B and C). The residual image, which contained the low surface brightness emission, was then deconvolved with the maximum entropy routine \textsc{vtess} with a starting model of the central part of the residual image smoothed with a circular Gaussian of full width at half maximum (FWHM) of 2 arcsec. Finally, the high surface brightness features were restored with the AIPS task \textsc{rstor} into the low surface brightness map and the combined total intensity map was produced.

\begin{figure}
\begin{center}
\includegraphics[height=8cm,width=8cm]{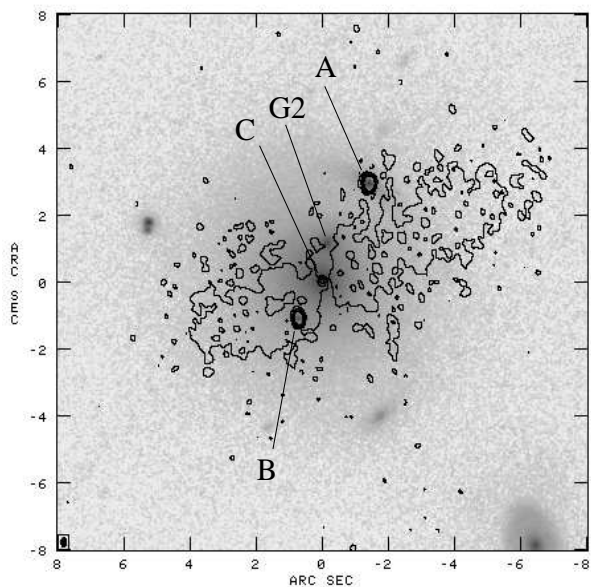}
\caption{{\it HST I}-band image (F814W) of B2108+213 in grey scale overlaid with the contours at 1.4~GHz using MERLIN. The lensed images A and B are coincident in the optical and the radio. The core component C coincident with galaxy G1 shows low surface brightness emission extending on either side in the radio. Galaxy G2 does not show any counterpart in the radio. The contours are ($-$3, 3, 6, 12, 24, 48, 96, 192)$\times$0.03~mJy~beam$^{-1}$ (rms in the map). North is up and east is to the left.}
\label{merjet}
\end{center}
\end{figure}

Fig. \ref{merjet} shows the total intensity map made from the combined datasets. Components A, B and C in this image are restored with a 0.252$\times$0.165 arcsec$^2$ beam. For the first time we find extended emission on either side of component C spreading over an area of $\sim$10$\times$2 arcsec$^2$.  It appears as though the extended jet emission originates from component C. The positions and the flux densities obtained from fitting Gaussian model components to A, B and C with the task \textsc{jmfit} in AIPS are given in Table \ref{mer1.4}. The total flux-density for the extended structure, measured by integrating over the region within the 3-$\sigma$ limit, is $\sim$ 70~mJy. Earlier measurements from the NVSS\footnote{National Radio Astronomy Observatory Very Large Array Sky Survey} catalogue \citep{condon98} suggest a flux density of $\sim$ 53~mJy for the system. The map of linear polarized emission shows only image A, with a polarized flux density of 0.6$\pm$0.1~mJy.  The rms noise of the polarized image is $\sim$0.1~mJy~beam$^{-1}$. Therefore, these observations were not sensitive enough to detect any polarized emission from image B, assuming that the polarized flux ratio between A and B is same as the flux ratio.

\begin{table}
\caption{The flux densities and positions of A, B and C from fitting Gaussian model components to the MERLIN 1.4~GHz map.}
\label{mer1.4}
\begin{tabular}{c r r c c}
Comp.	& RA			& Dec 			& $S_{peak}$		&$S_{total}$\\
	& (mas)			&(mas) 			& (mJy~beam$^{-1}$) 	& (mJy)\\
\hline 
A	& 0.0$\pm$0.2		& 0.0$\pm$0.2 		& 16.3$\pm$0.8 		&16.4$\pm$0.8\\
B	& 2141.3$\pm$0.4 	& $-$4026.7$\pm$0.4 	& 7.1$\pm$0.5 		& 7.7$\pm$0.4\\
C 	& 1434.4$\pm$6.0 	& $-$2915.6$\pm$6.0 	& 0.5$\pm$0.1 		& 0.6$\pm$0.1\\
\hline
\end{tabular}
\end{table}

\subsection{VLBA 1.7~GHz observations}

\begin{figure*}
\includegraphics[height=12cm,width=18cm]{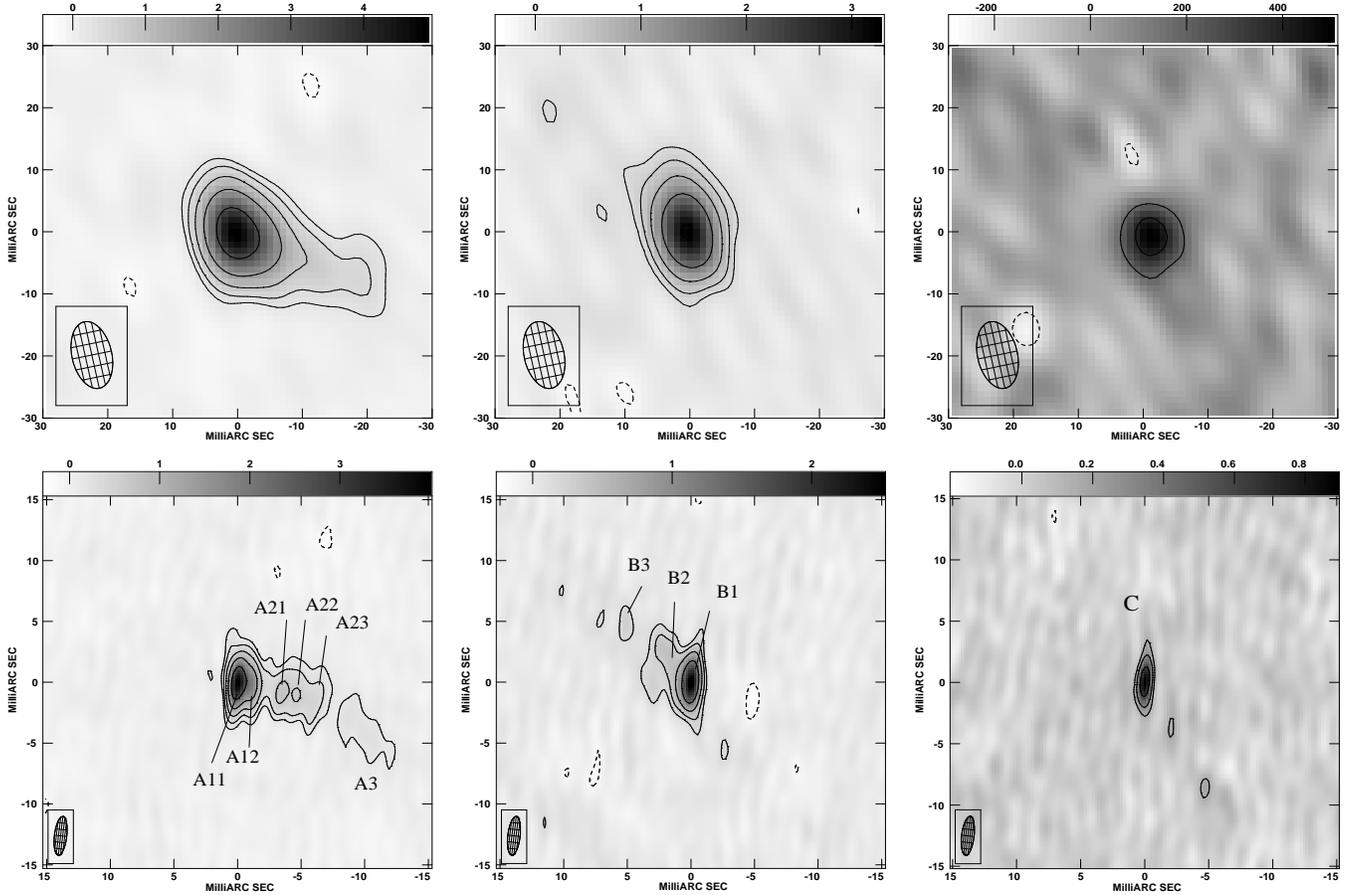}
\caption{The top panel shows VLBA 1.7~GHz maps of B2108+213 restored with a beam size of 10.9$\times$6.1~mas$^{2}$ and position angle of $12.18^{\circ}$. The bottom panel shows Global VLBI 5~GHz maps restored with a beam size of 3.19$\times$0.97~mas$^{2}$ and position angle of $-6.17^{\circ}$. Component A shows extended structure to the south-west direction at both frequencies. Component B shows north-east extension, which is better resolved at 5~GHz.  Component C which is coincident with the main lensing galaxy, is compact at both frequencies. The contours for all of the maps are ($-$3, 3, 6, 12, 24, 48)~$\times~\sigma_{map}$ given in Table \ref{tab18} for the 1.7~GHz dataset and is 0.05~mJy~beam$^{-1}$ for the 5~GHz dataset. North is up and east is left. Grey-scales are in mJy~beam$^{-1}$, except for the 1.7~GHz image of component C which is in $\mu$Jy~beam$^{-1}$.}
\label{maps18_6}
\end{figure*}

\begin{table*}
 \begin{center}
\caption{Positions, flux densities and rms noise of the components at 1.7~GHz.} 
\label{tab18}
\begin{tabular}{l r r c c r}
\hline
Comp.  & RA 		  & Dec			& $S_{peak}$	& $S_{total}$ 	& $\sigma$~$_{map}$ \\
       & (mas) 	  & (mas) 		& (mJy~beam$^{-1}$) & (mJy) 	& (mJy~beam$^{-1}$) \\
\hline 
A1     &     0.0$\pm$0.1 & 0.0$\pm$0.1		& 3.0$\pm$0.2	& 4.6$\pm$0.2 	& 0.070    \\
A2     &     3.0$\pm$0.2 & $-$0.9$\pm$0.2	& 2.3$\pm$0.1	& 2.4$\pm$0.1 	& 0.070	   \\
A3     & $-$12.2$\pm$0.5 & $-$5.5$\pm$0.5	& 0.6$\pm$0.1	& 1.7$\pm$0.1 	& 0.070	   \\
B      &  2135.9$\pm$0.1 & $-$4030.5$\pm$0.1	& 3.3$\pm$0.2	& 4.3$\pm$0.2 	& 0.065    \\
C      &  1429.0$\pm$0.5 & $-$2889.9$\pm$0.5	& 0.5$\pm$0.1	& 0.5$\pm$0.1 	& 0.065    \\
\hline
\end{tabular}
\end{center}
\end{table*}

B2108+213 was observed on 2002 June 19 with the VLBA at 1.7~GHz. The aim of this observation was to image any low surface brightness extended jet emission from the lensed images and to detect component C on mas-scales. The observing run lasted for 7.5~h in total, with $\sim$5.2~h spent observing B2108+213. The calibrator B2103+213 was used for phase-referencing, with a 5~min time cycle between the lens system (3.5~min) and the calibrator (1.5~min). The data were taken in the left-hand circular polarization through 4 IFs, each with 8~MHz bandwidth. The aggregate bit rate was 128 Mb\,s$^{-1}$ with 2 bit sampling. All of the VLBA antennas were available with the exception of Los Alamos. The data were correlated at the VLBA correlator where each IF was divided into 16$\times$0.5~MHz channels and averaged over 2-s time intervals. For wide-field imaging with interferometers, bandwidth smearing of the {\it uv}-data can cause radial smearing which reduces the peak flux density of a component far away from the phase centre. To limit the effect of bandwidth smearing, we correlated the data at two positions centred on images A and B. Note that for the set-up used here bandwidth smearing effects are significant\footnote{For a 10 percent loss in the measured peak flux density of a point source.} at a radius of $\ga$~11~arcsec, which is well outside the maximum lensed image separation (4.56~arcsec).

Both the A and B datasets were reduced individually using {\sc aips}. The data were amplitude calibrated from the system temperature and gain curve values for each antenna, and corrected for the change in the parallactic angle. The data for B2103+213 were phase and amplitude self-calibrated and the solutions were applied to the lens system B2108+213. The mapping of B2108+213 was carried out using {\sc imagr}, without further frequency or time averaging. We used an iterative process of cleaning and phase-only self-calibration, using a 9 min solution interval, to map the lens system.

In the upper panel of Fig. \ref{maps18_6}, we show the naturally weighted maps of lensed images A and B, and of radio component C from dataset B. Image A shows a compact core and a radio-jet extending to the south-west which was fitted by a three component Gaussian model. Previous VLBA imaging at 5~GHz by \citet{mckean05} found the jet to extend over 10 mas in scale, whereas the new deeper 1.7~GHz map shows more emission extending beyond 20 mas from the radio core. Image B shows a single core component with a hint of jet emission to the north-east. Image B is also fainter than image A, which is consistent with gravitational lensing where the surface brightness of the lensed images is conserved. A single component Gaussian fit successfully modelled the emission from image B. The radio component C was detected for the first time at mas-scales. A single component Gaussian fit shows that the emission from C is unresolved.  The results of fitting Gaussian model components using the AIPS task \textsc{jmfit} are listed in Table \ref{tab18}. Note that the extended emission detected in the MERLIN imaging has been resolved out here.

\subsection{Global VLBI 5~GHz observations}
\label{global5}
In order to better resolve the north-east extension in image B and to determine the spectral index of component C, we undertook a global VLBI observation of B2108+213 at 5~GHz. The lens system was observed on 2006 February 17 with the Effelsberg, Jodrell Bank Mk2, Westerbork, Medicina, Torun, Noto, Green Bank and the 10 VLBA antennas. The 14.5~h long observation was taken in both the right and left-hand circular polarizations, through 4 IFs each with 8~MHz bandwidth and a 256~Mb\,s$^{-1}$ bit rate. Here, the 5 min time cycle between the lens and the calibrator (B2103+213) was divided into 3 and 2 min scans, respectively for phase-referencing. The correlation was carried out at JIVE\footnote{Joint Institute for VLBI in Europe} where the data were divided into 16$\times$0.5~MHz channels and time-averaged over 1-s intervals. The maximum field-of-view defined by bandwidth and time-averaged smearing is $\sim$ 8~arcsec with this setup. Therefore, we obtained a single correlation of the data with the phase centre at the mid-point between lensed images A and B.

The data were reduced using {\sc aips} in a similar manner to the 1.7~GHz VLBA observations.  The Green Bank telescope had severe problems throughout the observing run, hence it was removed completely from the dataset. All baselines with Torun were discarded because of a poor amplitude calibration. Throughout the data reduction process all of the antennas were given equal weights to stop the large antennas (like the 100~m Effelsberg telescope) dominating the {\it uv}-dataset. This avoids higher side-lobes and a deterioration of the image quality. The data for B2108+213 were self-calibrated with a 3-min long solution interval, and \textsc{clean}ed using {\sc imagr}. The naturally weighted global VLBI maps of B2108+213 are presented in the lower panel of Fig. \ref{maps18_6}.

Image A shows the same core-jet structure observed at 1.7~GHz and previously at 5~GHz \citep{mckean05}. These new Global VLBI data clearly resolve image A into three main components A1, A2 and A3. Component A1 is made up of two compact sub-components A11 and A12. Component A2 appears to be an extended knotty jet feature, which is further divided into three sub-components, A21, A22 and A23. The third component (A3) of image A is a faint jet feature which kinks towards the south. Components A1 and A2 were fitted with two and three elliptical Gaussian models respectively, using Powell's minimization method \citep{press92} in the image plane. Component A3 could not be well fitted with multiple Gaussian models. Therefore, the flux density was measured by summing all of the surface brightness emission within the 3-$\sigma$ boundary. The position of A3 was obtained from the surface brightness peak.  Note that due to the different angular resolutions and frequency dependent structure, the components (A1, A2 and A3) at 1.7~GHz do not correspond to those observed at 5~GHz -- this is simply our naming convention. The map of image B shows the expected core-jet structure to the north-east. We identify B1 as the core, and B2 and B3 as the counterparts to the jet features detected from the 5~GHz imaging of A. The emission from image B was fitted with three elliptical Gaussian components, using the same minimization method as for image A. Note that the small feature to the south of component B2 was excluded. Component C was fitted with a compact elliptical Gaussian model (deconvolved FWHM is 0.8 mas) at 5~GHz. There is no evidence of any collimated jet emission towards the extended lobes detected in the 1.4~GHz MERLIN observations.

The fitted positions and flux densities of the Gaussian model components are presented in Table \ref{all6}. The errors for the flux densities of A and B were calculated from the principle used in \cite{fom99}, and the errors on the positions were taken from the AIPS task \textsc{jmfit}.

\begin{table}
\caption{Positions and flux densities of the fitted Gaussian components for the 5~GHz data. Component A3 was not fitted with a two-dimensional Gaussian (see Section \ref{global5} for details).}
\label{all6}
\begin{tabular}{l r r c c}
\hline
Comp.	& RA	& Dec	& $S_{peak}$		& $S_{total}$ \\
       		& (mas)	& (mas)	& (mJy~beam$^{-1}$)	& (mJy)	\\
\hline 
A11	& 0$\pm$0.1 		& 0$\pm$0.1	    	& 3.6$\pm$0.4	& 3.8$\pm$0.5 	\\
A12	& $-$1.1$\pm$0.1 	& 0.0$\pm$0.1		& 1.9$\pm$0.3 	& 2.6$\pm$0.3 	\\
A21	& $-$3.5$\pm$0.1 	& $-$0.6$\pm$0.1	& 0.7$\pm$0.2 	& 1.2$\pm$0.2 	\\
A22	& $-$4.7$\pm$0.1 	& $-$0.5$\pm$0.1	& 0.3$\pm$0.1 	& 0.3$\pm$0.1 	\\
A23	& $-$6.0$\pm$0.1 	& $-$1.3$\pm$0.1	& 0.5$\pm$0.2 	& 1.4$\pm$0.2 	\\
A3	& $-$9.8$\pm$0.5 	& $-$3.9$\pm$0.5	& 0.4$\pm$0.2 	& 0.4$\pm$0.2 	\\
B1	& 2135.2$\pm$0.1 	& $-$4030.0$\pm$0.1	& 2.3$\pm$0.3 	& 2.9$\pm$0.4 	\\
B2	& 2137.0$\pm$0.3 	& $-$4028.0$\pm$0.3	& 0.5$\pm$0.2 	& 1.1$\pm$0.2 	\\
B3	& 2140.5$\pm$0.4 	& $-$4025.2$\pm$0.4	& 0.3$\pm$0.1 	& 0.3$\pm$0.1 	\\
C	& 1430.7$\pm$0.1	& $-$2888.1$\pm$0.1	& 0.9$\pm$0.1 	& 0.9$\pm$0.1 	\\
\hline
\end{tabular}
\end{table}

\section{Mass Model}
\label{model}

We now investigate mass models for the B2108+213 lens system. The current best model has two singular isothermal spheres for galaxies G1 and G2, and an external shear \citep{mckean05}. This model reproduces the observed positions and flux densities of the lensed images, but has no degrees of freedom due to the limited number of observational constraints. Using the new high resolution VLBI observations presented here, we first test whether the image configuration can be better explained by a single lensing galaxy (G1), or by two galaxies (G1 and G2). We then test whether component C can be a core lensed image. Finally, we investigate the influence of the environment on the mass model. The lens models were generated using the publicly available code {\sc lensmodel} \citep{keeton01}.

\subsection{Case 1: C as the active nucleus of lensing galaxy G1}
\label{case1}

The position of radio component C was used as the position for the lens galaxy G1 in this case. We note that the position obtained for component C from the VLBI imaging is not consistent with the position measured for G1 from the {\it I}- and {\it H}-band imaging (F814W and F160W, respectively) with the {\it Hubble Space Telescope}. This is almost certainly due to G1 having a complicated surface brightness profile, particularly in the infrared (this will be discussed further by McKean et al. in prep.). However, our position for component C is consistent with the {\it V}-band (F555W) imaging. Therefore, the position of G2 and the flux ratio between G1 and G2, which are used in the models, are taken from the {\it V}-band data (see Table 4 of \citealt{mckean05}).

As expected from lensing, we believe that the three main components of image A are the counterparts of the components in image B.  However, image A is further resolved and shows several sub-components in A1 and A2. The compact sub-component A11 is probably the core, hence it was used as the position for A1. As component A2 is much more extended, a flux-density weighted position was determined from the sub-components ($\Delta \alpha=-$3.8$\pm$0.2~mas, $\Delta \delta=-$0.7$\pm$0.2~mas relative to A11). For component A3 and all of the components of lensed image B, the positions from Table \ref{all6} were used. The flux densities for only A1 (i.e. A11+A12) and B1 were used for the modelling. The total number of constraints to the lens models provided from the lensed images is 13.

First, we consider the lens as a single galaxy G1, with a singular isothermal sphere (SIS) density profile, with an external shear. This model has 9 parameters (1 for the Einstein radius, 2 for the shear and position angle, and 6 parameters for the three source positions) and 4 degrees of freedom (dof). The total $\chi^2_{4}$ of the model is 7.6, where the subscript refers to the dof, and the results are given in Table \ref{mod1}. This simple model fits the data well, and requires a large Einstein radius of 2.18~arcsec and an external shear of 0.05 at a position angle of 107.3~deg.

We now test what contribution G2 has if it is included as part of the lens potential.  Hence, we introduce galaxy G2 as a companion lensing galaxy at the same redshift as G1.  A SIS centred on the optical position of G2 was added to the SIS+shear model for G1.  Following \citet{mckean05}, the mass ratio between G1 and G2 was fixed to $b_{G2}/b_{G1}=$~0.29$\pm$0.05, from the optical luminosities.  Note that including G2 in this way does not change the number of dof. The total $\chi^2_{4}$ of the model is 6.9 which is only marginally better than the single SIS+shear model used for G1. The fitted parameters are given in Table \ref{mod1}. Including G2 lowers the Einstein radius of G1 to 1.71~arcsec and reduces the shear to 0.04.

\begin{figure}
\includegraphics[height=14cm,width=7cm]{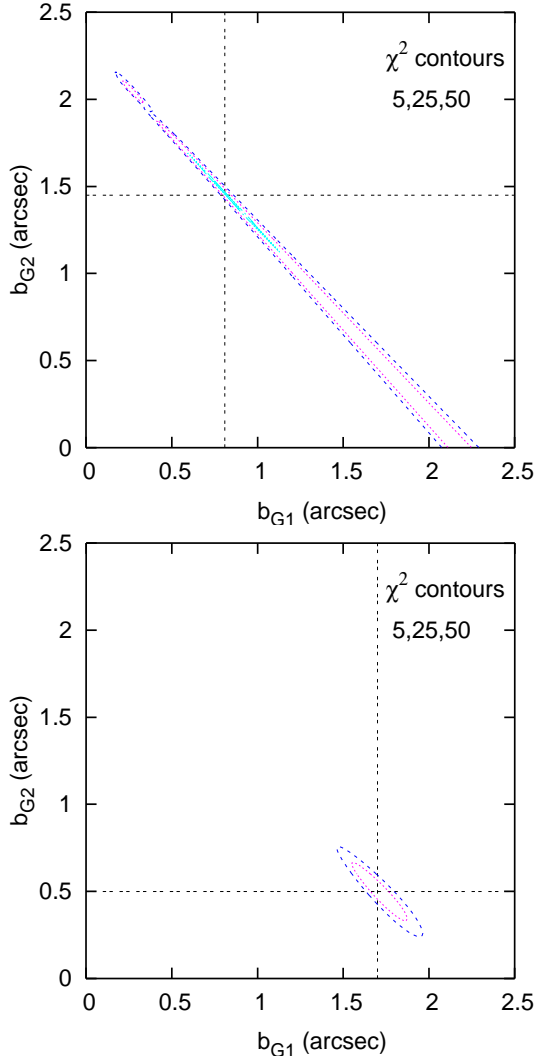}
\caption{A degeneracy between the Einstein radii of G1 and G2 for the
SIS+SIS+shear model is shown in the upper panel. Without any
constraint on the masses of the galaxies, the best-fitting model requires
G2 to be four times more massive than G1, which contradicts the
optical luminosities. Introducing a constraint on the mass ratio
between G1 and G2 breaks this degeneracy as shown in the lower
panel. The intersection point of the double-dotted straight lines
marks the minimum in both plots.}
\label{bg12}
\end{figure}

Note that if the ratio of the masses is not fixed and the Einstein radius of G2 is left as a free parameter, then the Einstein radius of G1 is lowered further to 0.76~arcsec and the shear is reduced to 0.01. Also, the Einstein radius of G2 increases to 1.5~arcsec. Here, the simplest model that can be fitted needs to have the main lensing galaxy positioned collinearly between the two lensed images. Since G2 is closest to this, it is taken as the main perturber and the shear is reduced. The reduced $\chi^2$ of this model is 0.9, which is lower than for the case when the mass ratio is fixed (reduced $\chi^2=$~1.7) and for the single galaxy model (reduced $\chi^2=$~1.9). In Fig. \ref{bg12}, we show the $\chi^2$ plot of the Einstein radii of G1 and G2. There is a clear degeneracy between the Einstein radii of G1 and G2. However, since G1 is clearly the dominant mass clump of the system as shown from the optical data, the prior from the optical luminosity is used for the SIS+SIS+shear model to break this degeneracy and to produce a physically plausible model.

So far, only an external shear was used to account for any mass ellipticity in the model. It is clear that some ellipticity is required to account for the non-collinearity between the lensed images and lensing galaxy G1. The ellipticity in the model could be due to G1 having an elliptical halo. Hence, the external shear was replaced with an elliptical mass distribution for G1, while an SIS is fixed at the position of G2. The dof of this singular isothermal ellipsoid (SIE)+SIS model is 4, and the resulting reduced $\chi^2$ is 1.6. This model fits the image positions slightly better than previous models without changing the fitted parameters significantly (see Table \ref{mod1}). The ellipticity of G1 is found to be 0.135 at a position angle of 105.5~deg. The ellipticity of the surface brightness profile of G1, as measured from the {\it HST} data, is 0.14 at a position angle of 57~deg \citep{mckean05}. The offset between the position angle of the halo and the light distribution of G1 suggests that the environment may also be affecting the shape of the lensing potential. However, since there are only two lensed images of B2108+213 it is not possible to constrain both an elliptical potential for G1 and an external shear. Moreover, since ellipticity and shear (and also the presence of another lensing galaxy, G2) will produce similar observable effects, the actual origin of the asymmetry between the lensed images and the lensing galaxy G1 is difficult to determine.

\begin{table}
\begin{center}
\caption{The fitted parameters with the respective 1-$\sigma$ uncertainties for the single lensing galaxy case (SIS+shear), and the two lensing galaxies cases (SIS+SIS+shear and SIE+SIS) with the mass ratio of G2 and G1 fixed. The angular separations relative to image A1 and the Einstein radii are in milliarcsecond. The position angle of the shear and ellipticity is measured in degrees east of north. The superscript `*' refers to the values resulting from the fitted models.}
\label{mod1}
\begin{tabular}{l r r r}
\hline
Name & \multicolumn{3}{c}{Fitted values}\\ 
                                & SIS+shear		& SIS+SIS+shear         & SIE+SIS           \\ \hline
G1 ($\alpha$, $\delta$)         & 1431$\pm$1	        & 1431$\pm$1            & 1431$\pm$1    \\
            			& $-$2888$\pm$1	        & $-$2888$\pm$1         & $-$2888$\pm$1 \\
G2 ($\alpha$, $\delta$) 	&		        & 1273$\pm$5            & 1273$\pm$5    \\
            			& 			& $-$1786$\pm$5         & $-$1786$\pm$5 \\
G1 Einstein 	        	& 2177$\pm$16.0         & 1711$\pm$38           & 1736$\pm$38       \\
radius                          &                       &                       &                   \\
G2 Einstein      	        &       	        & 498$\pm$39            & 506$\pm$39        \\
radius                          &                       &                       &                    \\
External Shear          	& 0.050$\pm$0.007	& 0.038$\pm$0.005       &                    \\
Position angle          	& 107.3$^{+1.3}_{-1.8}$ & 105.6$^{+1.5}_{-2.0}$ &                     \\
Ellipticity                     &                       &                       & 0.135$\pm$0.017       \\
Position angle          	&                       &                       & 105.5$^{+1.6}_{-2.1}$  \\
%                                &                       &                       &                   \\
Source 1 ($\alpha$, $\delta$)$^*$	& 989.4		    	& 1059.0                & 1057.6           \\
                                        &  $-$2110.3		& $-$2058.8             & $-$2021.6        \\
Source 2 ($\alpha$, $\delta$)$^*$	& 987.7			& 1057.6                & 1056.2           \\
			                &  $-$2110.1		& $-$2058.3             & $-$2021.1        \\
Source 3 ($\alpha$, $\delta$)$^*$	& 985.4			& 1055.7                & 1054.4           \\
         		        	&  $-$2111.3		& $-$2059.1             & $-$2021.7        \\
$S_{BA}$ ($S_{B1}/S_{A1}$)$^*$     	& 0.64 		        & 0.63                  & 0.63             \\
\hline
\end{tabular}
\end{center}
\end{table}

\begin{figure}
 \includegraphics[height=8cm,width=8cm]{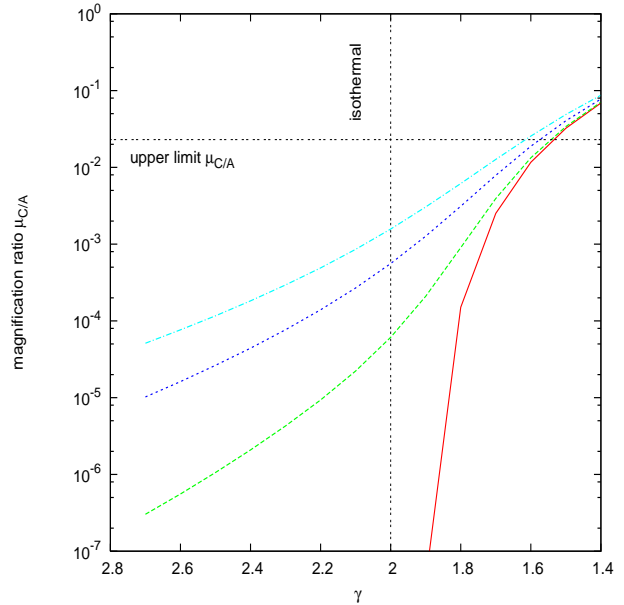}
\caption{The relative magnification of image A and a possible core lensed image C as a function of density power-law slope for a single lensing galaxy (G1) in the absence of a supermassive black hole. The solid (red) line is for a core radius of 0.005~pc. The dashed (green), dotted (blue) and dashed-dotted (cyan) lines are for core radii of 50, 150 and 250~pc, respectively. An upper limit to the relative magnification between images A and C is shown.}
\label{slcr}
\end{figure}

\begin{figure}
 \includegraphics[height=8cm,width=8cm]{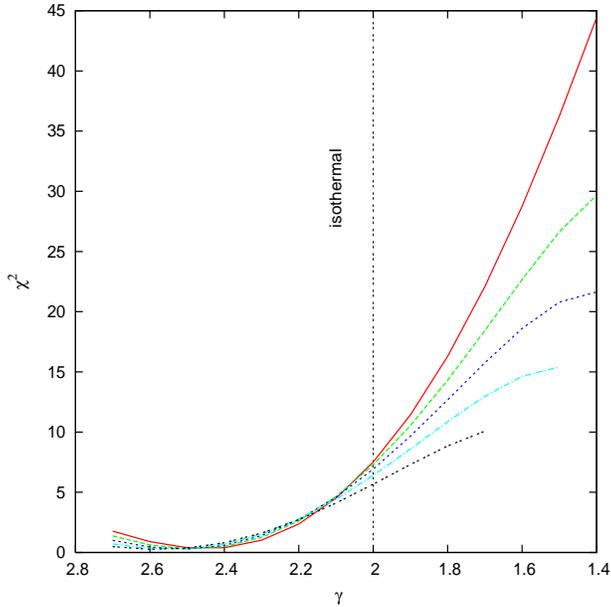}
\caption{The $\chi^2$ as a function of density power-law slope for the main lensing galaxy (G1). The solid (red) line is without the inclusion of galaxy G2. The dashed (green), dotted (blue), dashed-dotted (cyan) and double-dotted (black) lines are for models that include G2 as an SIS with an Einstein radius of 0.3, 0.5, 0.7 and 0.9~arcsec.}
\label{sl2gal}
\end{figure}

\begin{figure}
 \includegraphics[height=8cm,width=8cm]{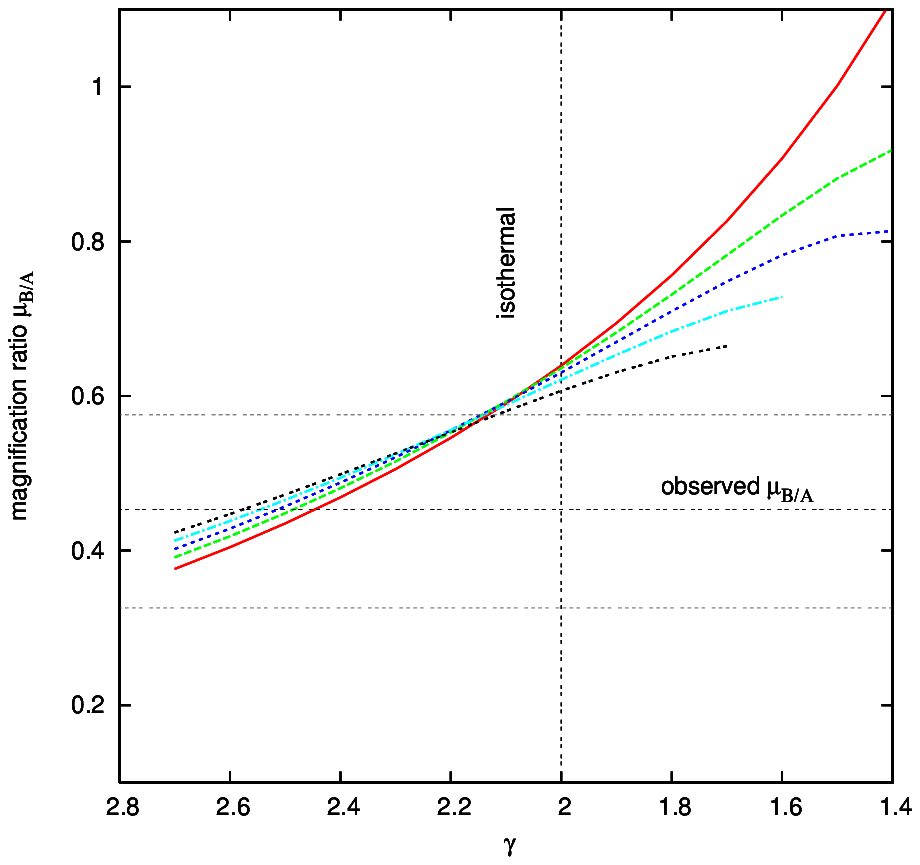}
\caption{The relative magnification between images A and B as a function of density power-law slope for a single lensing galaxy (G1). The solid (red) line is without the inclusion of galaxy G2. The dashed (green), dotted (blue), dashed-dotted (cyan) and double-dotted (black) lines are for models that include G2 as an SIS with an Einstein radius of 0.3, 0.5, 0.7 and 0.9~arcsec. The observed flux ratio at 5~GHz is shown with a conservative 20 per cent uncertainty in the flux densities of images A and B.}
\label{slmag}
\end{figure}

\subsection{Case 2: C as the core lensed image}
  
Extended mass distributions are expected to produce an extra lensed image near the position of the lensing galaxy (\citealt{dyrod80}; \citealt{burke81}). However, the mass density distribution close to the centre of the lensing galaxy will affect the magnification of the core lensed image. For example, an isothermal density profile will completely demagnify the core lensed image. Since galaxies are known to have global density profiles close to isothermal (e.g. \citealt{koopmans06}), searches for core lensed images tend to concentrate on asymmetric double image systems (e.g.  \citealt{boyce06}; \citealt{zhang07}), that is, lens systems where the A:B flux ratio is greater than 10:1. This is because in these cases the magnification of the core lensed image is highest and these systems offer the best possible chances of detection. The only example of a known galaxy-scale lens system with a core lensed image is PMN~J1632$-$0033 \citep{winn03b}, which is an asymmetric double with a flux ratio of 15:1. For B2108+213, the flux ratio of image A to B is about 2:1, so it seems unlikely that component C could be a core lensed image. However, given the possibility that the lens potential may be made up of several galaxies, the overall density profile might be shallower than isothermal. Therefore, we now explore different mass models to establish the nature of component C.

Throughout this section the position of galaxies G1 and G2 are taken from the {\it V}-band data \citep{mckean05} and the mass ratio is fixed. The positions and flux densities of A, B and C are taken from the global VLBI 5~GHz data (Table \ref{all6}). For the isothermal models presented in Section \ref{case1}, no core lensed image is produced. Therefore, we investigate mass distributions for G1 that deviate from isothermal close to the centre of the lensing galaxy, that is, by inserting a core radius at the inner part of the halo.

We use a Non-singular Isothermal Sphere (NIS) halo for G1 which has an isothermal mass profile with a flat density core. The second lensing galaxy G2 is too far away for its core properties to significantly affect the magnification of the central image. Therefore, we choose a SIS profile for galaxy G2. An external shear is also included in the model. This new model has three additional constraints from the position and flux density of component C and 10 free parameters to fit, thereby giving 6 dof. The model fits the position of component C as a core lensed image well, but cannot fit the flux density, resulting in a very high total $\chi^2_{6}$ of 88.6. The best-fitting core radius is 0.1~mas ($\equiv$~0.5~pc) and gives a flux density of $10^{-6}$~mJy for the core lensed image, which is still several orders of magnitude fainter than the observed flux density of component C. This model fails because a much higher core radius is required to fit the flux density of C, but this will be at the expense of fitting the position of component C and the positions and flux densities of the lensed images A and B. Therefore, it seems very unlikely that the third radio component is a core lensed image because the observed flux-density is just too high.

\subsection{Limits on the density profile of G1}

For dark matter dominated structures like galaxy clusters, the inner density profile of the mass distribution tends to be quite shallow (i.e. $\gamma \sim$1--1.5). This is based on the results from numerical simulations (e.g. \citealt{navarro96,moore98}) and from gravitational lensing (e.g. \citealt{sand05}). In this section, we attempt to place constraints on the density profile of the lensing halo, with the aim of determining whether it is consistent with an isothermal mass profile, or a shallower dark-matter dominated profile. To do this, we assume that the lensed image is not coincident with component C. We then derive an upper limit to the flux-density of a core lensed image using the 3-$\sigma$ flux-density limit determined from the rms map noise ($S_C \leq$~150~$\mu$Jy). 

We introduce a group halo with a power-law density profile, $\rho \propto r^{-\gamma}$, centred on the position of G1. In addition, we incorporate a core radius of 0.005, 50, 150 and 250~pc. The model includes an external shear. The VLBI 5~GHz data for lensed images A and B are used to constrain the model. The Einstein radius, external shear and position angle are then optimized for density profiles with $\gamma$ between 1.4 and 2.7, and for each value of the core radius. In Fig. \ref{slcr}, we show the relative magnification between lensed image A and a possible core lensed image C as a function of power-law density profile slope. We have also marked the upper limit to the relative flux ratio of a possible core lensed image and image A. First, it is clear that steeper density profiles result in a more demagnified core lensed image, as expected. Also, increasing the core radius from 0.005 to 250~pc, increases the magnification of the core lensed image. Interestingly, the upper limit to the relative magnification requires that the density profile of any group halo be steeper than $\sim$1.5.

The presence of a radio-loud AGN implies that there is also a supermassive black hole at the centre of the lensing galaxy G1. The presence of a black hole is expected to increase the number of core lensed images to two, or possibly demagnify a core lensed image completely, depending on the mass of the black hole \citep*{mao01,rusin05}. Using the stellar velocity dispersion of the lensing galaxy G1 ($\sigma_v =$~360~km\,s$^{-1}$; McKean et al. in preparation) and the known correlation between black hole mass and the stellar velocity dispersion of the host galaxy, we find that the black hole associated with the AGN within G1 should have a mass of $\sim$10$^{9}$~M$_{\odot}$. Including this black hole as a point mass at the centre of the G1 halo results in any core lensed image being completely suppressed.

Our mass models which have G1 as the only lensing galaxy (SIS), or also include G2 as a companion lensing galaxy (SIS) to G1 (SIS or SIE), fitted the positions of the lensed images well, but failed to recover the flux ratio; the observed flux ratio is $S_B/S_A =$~0.45 at 5~GHz whereas the modelled flux ratio is $\sim$0.63 (see Table \ref{mod1}). This may suggest that the overall density profile of the lens system differs from isothermal, possibly due to the surrounding environment. Therefore, we now investigate whether a different density profile is a better fit to the data. For this model, we again use a variable power-law density profile for G1. As we are no longer using an isothermal mass profile for G1, it is not possible to fix the ratio of the Einstein radii of G1 and G2 from their optical luminosities via the Faber--Jackson relation. Therefore, we include G2 as a singular isothermal sphere with a fixed Einstein radius of 0, 0.3, 0.5, 0.7 and 0.9 arcsec to test different scenarios. An external shear is also included. This model has 9 parameters and 13 constraints. In Fig. \ref{sl2gal}, we show the model $\chi^2$ as a function of power-law density slope and for different Einstein radii for G2. It is clear that those models with a steeper density profile are a better fit to the data. However, increasing the Einstein radius of G2 also improves the fit, but not to the extent as to justify shallower mass profiles for G1. Note that for the cases where the Einstein radius of G2 is 0.7 and 0.9 arcsec, shallow mass profiles for G1 (i.e. $\gamma \leq$ 1.7 and 1.5, respectively) produce four lensed images due to the introduction of a new minimum and saddle point in the time-delay surface. Also, the flux density of image A is lower than that of image B. Therefore, these models can be ruled out. In Fig. \ref{slmag}, we show the predicted flux ratio of images A and B as a function of the density profile of G1 and for the same set of Einstein radii for G2. The observed flux ratio at 5~GHz, with a 20 per cent error on the flux densities of images A and B, is also shown. It is clear that shallow density profiles are inconsistent with the flux ratio of A and B, if the halo is fixed at the position of G1 and assuming the flux densities are not significantly affected due to variability and a time delay. Furthermore, models with shallower profiles (i.e. $\gamma \leq$ 1.7) require the flux density of B to be close to that of A, which is not consistent with the observed flux ratio. The best-fitting models require the density slope for G1 to be steeper than isothermal ($\gamma=2.45_{-0.18}^{+0.19}$, for a single spherical mass model), and are fairly insensitive to the Einstein radius of G2.

\section{Discussion}
\label{disc}

The aims of this paper were to determine the nature of the third radio component and to investigate the B2108+213 lensing potential, with particular emphasis on the contribution of the second lensing galaxy and the group environment. We now discuss to what extent these aims have been met.

\begin{figure}
\begin{center}
 \includegraphics[height=8.5cm,width=8.5cm,angle=-90]{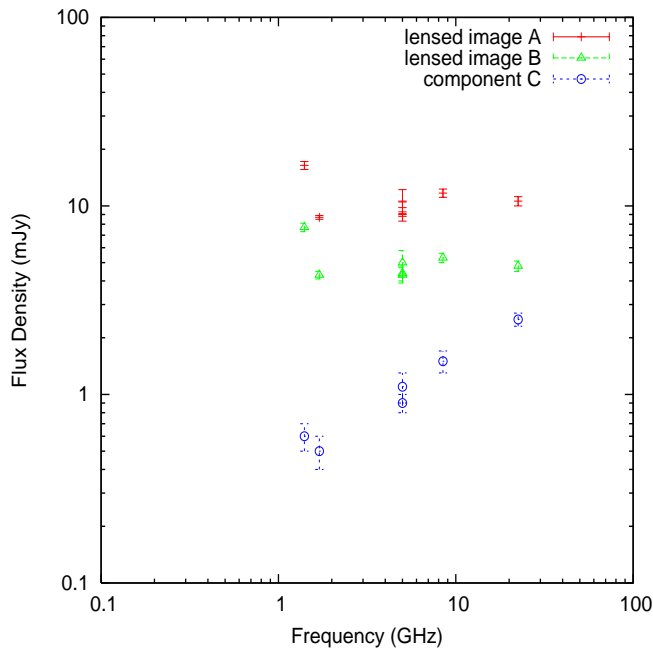}
\caption{The radio spectra of lensed images A and B, and component C. The 1.4, 1.7 and 5~GHz data points are from this paper. The rest of the data points are taken from \citet{mckean05}.}
\label{spec}
\end{center}
\end{figure}

\subsection{The lensing potential}

The new 1.7 and 5~GHz data presented here have found compact and extended emission from the two lensed images A and B which is consistent with gravitational lensing. We find that the surface brightnesses of the two lensed images are as expected for gravitational lensing. Furthermore, the series of three non-collinear components detected in the 5~GHz imaging of images A and B appear to have mirror symmetry. Hence, the expected parity reversal of the lensed images is also observed. In Fig. \ref{spec} we show the radio spectral energy distributions of images A and B. These spectra have been constructed using the data presented here and from the data presented by \citet{mckean05}. It is clear that both lensed images have very similar flat radio spectra. We note that the MERLIN 1.4~GHz flux-densities are much higher than the VLBA 1.7~GHz flux-densities for images A and B. This is either due to calibration errors in the MERLIN or VLBA data, or more likely, that extended jet emission from the two lensed images has been resolved out by the high resolution 1.7~GHz imaging.

We find that isothermal mass models which have only G1 as the lensing galaxy or which include the companion lensing galaxy G2, are a good fit to the data. However, the two lens galaxy model appears to be a slightly better fit. Both of these lens models fit the positions of the lensed images well, but do not fit the flux ratio well. This could be due to variability or a time delay in the radio-loud lensed images, or may indicate that a more complex mass model including additional group galaxies is required. Although B2108+213 has not been monitored for variability there are 3 independent observations carried out with MERLIN and the VLBA at 5~GHz to a low surface-brightness limit. As can be seen from Fig. \ref{spec} the flux densities from these observations are consistent to within their uncertainties. It seems likely that the lens environment is playing a role in the image splitting of B2108+213. We find that dark matter dominated haloes, that is, those with shallow density profiles ($\gamma \leq$~1.5) seem not to be consistent with the observed flux ratios of the lensed images, or the non-detection of a core lensed image (in the absence of a supermassive black hole). In fact, density profiles that are steeper than isothermal are preferred. Even though the density profiles of isolated galaxies appear to be consistent with isothermal (e.g. \citealt{koopmans06}), it is predicted that galaxies undergoing an interaction can have density profiles that are steeper than isothermal for $\la$~0.5~Gyr after the initial interaction, before returning to the original isothermal state \citep*{dobke07}. Given the close proximity of galaxy G2, this scenario is certainly consistent with the lens system B2108+213. However, further observations will need to be carried out to confirm that the density profile of G1 is steeper than isothermal. In particular, measuring the redshift and stellar velocity dispersion of the companion galaxy G2, coupled with the source redshift, will allow the inner density profile of the lensing mass distribution to be determined. Also, if the background source yields a time-delay in the future, then the density profile can be found by assuming the Hubble constant (e.g. \citealt*{dobke06}; \citealt{auger07b}; \citealt*{read07}). Since all of the mass models tested here require an external shear of $\sim$4--5 per cent, it is likely that additional group/cluster galaxies are contributing to the overall potential. The impact of the group/cluster galaxies associated with G1 will be discussed in a follow-up paper.

\subsection{A radio-loud lensing galaxy}

It seems almost certainly the case, for several reasons, that the compact emission from radio component C is due to a radio-loud AGN hosted within the main lensing galaxy G1. First, as can be clearly seen from Fig. \ref{spec}, the radio spectrum of component C is rising from 1.4 to 22.46~GHz, whereas the radio spectra of the lensed images A and B are flat. The spectral index of component C is $\alpha_{1.4}^{22.46} =$~0.51$\pm$0.01. The spectral difference between component C and the two lensed images could be due to free-free absorption from the interstellar medium of the lensing galaxy (e.g. as in the case of PMN\,1632-0033; \citealt{winn03b}). However, the effect of free-free absorption is known to be a strong function of frequency, and the spectrum of component C shows no curvature and is constantly rising from 1.4 to 22.46~GHz. Second, the radio emission is coincident with the optical position of the lensing galaxy G1, as expected for an AGN. Third, although it is possible to fit a core lensed image at the position of component C, the flux density of component C is much too large to be a third lensed image. Note that this would be exacerbated if component C has been affected by free-free absorption.

The classification of component C as the core of an AGN within G1 is also consistent with the extended lobe emission on either side of the lensing galaxy detected by MERLIN at 1.4~GHz. Moreover, the diffuse lobed morphology of the extended emission is as expected for a Fanaroff-Riley type I radio source (FR\,I; \citealt{fanril74}). There are no highly collimated jets or hotspots which are typically seen in the more powerful FR\,II sources. The total 1.4~GHz rest-frame luminosity of the radio emission associated with G1 is $L_{1.4} \sim$~10$^{25}$~W~Hz$^{-1}$ (this assumes a spectral index of $\alpha =-$0.8 for the lobe emission). Extrapolating this luminosity to 178~MHz gives $L_{0.178} \sim$~10$^{24}$~W~Hz$^{-1}$~sr$^{-1}$, which is below the FR\,I--FR\,II luminosity divide of $L_{0.178} \sim$~10$^{25}$~W~Hz$^{-1}$~sr$^{-1}$ \citep{fanril74}. Finally, the absolute {\it R}-band magnitude of the lensing galaxy G1 is $M_R =-$24. From the correlation between the rest-frame 1.4~GHz luminosity and the absolute {\it R}-band magnitude of the host galaxy (see fig. 1 of \citealt{ledowe96}), which divides sources into FR\,I and FR\,II, we find that the G1 radio source is in the FR\,I region. Although component C appears to be unresolved in the 1.4 GHz MERLIN and 1.7~GHz VLBA imaging, the presence of the extended jet emission may explain why component C was slightly extended in the MERLIN 5~GHz map presented by \citet{mckean05}. 

There are now two gravitational lens systems from the CLASS survey with a known radio-loud lensing galaxy; the other is CLASS B2045+265 \citep{fassnacht99,mckean07}. This gives a fraction of 11$\pm$8 per cent for lens galaxies with a radio-loud AGN from CLASS. This agrees closely with the percentage found from deep radio imaging of optically selected gravitational lens candidates from the Sloan Digital Sky Survey ($\sim$10 per cent; \citealt{boyce06b}).

It is possible to probe different lines-of-sight by searching for differences in the properties of lensed images, which should be identical in the absence of variability. Since image separations are typically only $\sim$0.5--1.5~arcsec, it is the lensing galaxy which is mostly being probed. This technique has been most successful at optical wavelengths where the dust extinction along the lines-of-sight to each of the lensed images has been used to test galactic extinction laws in high redshift lens galaxies \citep{falco99,wucknitz03,ardis06}. At radio wavelengths, free-free absorption of one (or more) of the lensed images has probed the ISM of the lensing galaxy (e.g. \citealt{winn03a}; \citealt{mittal07}; \citealt{wiklind96}). In the case of B2108+213, we have the unique situation of a radio jet from an AGN within the lensing galaxy passing in front of one of the lensed images (see Fig. \ref{merjet}). Therefore, B2108+213 can be used to probe the composition of an FR\,I radio jet. As can be seen from Fig. \ref{spec}, the flux ratio between images A and B is almost constant from 1.4 to 8.46~GHz\footnote{The 22.46~GHz values have been determined by assuming the flux ratio is unchanged at 22.46~GHz - see \citet{mckean05} for details.}. As such, there is no evidence of free-free absorption of the flux density of image B. Another possible propagation effect is depolarization/Faraday rotation of the emission from lensed image B as it passes through the magnetized plasma of the radio-jet. However, the MERLIN data presented here were not deep enough to establish whether or not image B has been depolarized. We do know that the background source is polarized by $\sim$4 per cent at 1.4~GHz from the data obtained for lensed image A. Therefore, further radio imaging should be carried out to determine if the properties of image B have been affected by the radio jet.

\section{Conclusions}
\label{conc}

We have presented new high resolution MERLIN and VLBI imaging of the gravitational lens system B2108+213, whose wide image separation is consistent with a massive lensing galaxy or possibly a group of galaxies. The VLBI imaging at 1.7 and 5~GHz found extended emission in the lensed images whose surface brightness and parities are consistent with gravitational lensing. Using the new constraints provided from the two lensed images we have tested mass models for the B2108+213 lens potential. We find that the properties of the lensed images are consistent with either a single massive lensing galaxy, or a two galaxy lens model which accounts for a nearby companion to the main lensing galaxy. In these cases the companion galaxy G2 was always represented by a singular isothermal sphere whereas the main galaxy was modelled as both an isothermal sphere and ellipsoid. We find that steeper than isothermal density profiles for the main lensing galaxy are a better fit to the data. Further models which include additional group galaxies will be tested when a spectroscopic survey of the B2108+213 local environment has been completed. 

Emission from the third radio component, which is coincident with the optical position of the main lensing galaxy, was found for the first time at mas-scales at both 1.7 and 5~GHz. Furthermore, MERLIN imaging at 1.4~GHz detected extended low surface brightness emission on either side of the third radio component. This jet emission has a morphology and luminosity which is consistent with an FR\,I type radio source. Attempts to model the radio core of the third radio component as a core lensed image failed because it was not possible to fit the observed flux-density (to within several orders of magnitude) by using a core radius and/or a variable power-law density profile for the main lensing galaxy. Therefore, we are confident that the third radio component is due to emission from an AGN embedded within the main lensing galaxy.

\section*{Acknowledgments}
AM would like to thank E. Boyce, C. Keeton, Y. Kovalev, D. Lal, S. More, P. Schneider and  O. Wucknitz for useful discussions. The VLBA is operated by the National Radio Astronomy Observatory which is a facility of the National Science Foundation operated under cooperative agreement by Associated Universities, Inc. The European VLBI Network is a joint facility of European, Chinese, South African and other radio astronomy institutes funded by their national research councils. MERLIN is a National Facility operated by the University of Manchester at Jodrell Bank Observatory on behalf of STFC. This work was supported by the European Community's Sixth Framework Marie Curie Research Training Network Programme, Contract No.  MRTN-CT-2004-505183 "ANGLES".

\label{lastpage}
\end{document}